\documentclass[a4paper]{jpconf}

\usepackage{enumerate}
\usepackage{amsmath,amssymb,bm}
\usepackage[dvips]{graphicx}
\usepackage{braket}   
\usepackage{url}   
\usepackage{color}
\usepackage{mathrsfs}

\begin{document}
\title{
Determination of mass of an isolated neutron star 
using continuous gravitational waves with two frequency modes: 
an effect of a misalignment angle 
}

\author{Kazunari Eda$^{1,2}$, Kenji Ono$^{1,3}$, and Yousuke Itoh$^2$}

\address{$^1$ 
Department of Physics, 
Graduate School of Science, 
University of Tokyo, Tokyo, 113-0033, Japan
}
\address{$^2$ 
Research center for the early universe, 
Graduate School of Science, 
University of Tokyo, Tokyo, 113-0033, Japan
}
\address{$^3$  
Institute for cosmic ray research, University of Tokyo, 5-1-5 Kashiwanoha, Kashiwa, Chiba 277-8582, Japan
}
 
\ead{eda@resceu.s.u-tokyo.ac.jp}
  
 \begin{abstract} 
  A rapidly spinning neutron star (NS) would emit a continuous gravitational wave (GW)
  detectable by the advanced LIGO, advanced Virgo, KAGRA and proposed third generation detectors 
  such as the Einstein Telescope (ET). 
  Such a GW does not propagate freely, 
  but is affected by the Coulomb-type gravitational field of the NS itself. 
  This effect appears as a phase shift in the GW depending on the NS mass.   
  We have shown that mass of an isolated NS can, in principle, be determined 
  if we could detect the continuous GW with two or more frequency modes. 
  Indeed, our Monte Carlo simulations have demonstrated that mass of a NS with 
  its ellipticity $10^{-6}$ at 1 kpc is typically measurable
  with precision of $20\%$ using the ET, if the NS is
  precessing or has a pinned superfluid core and emits GWs with once and twice the spin frequencies. 
  After briefly explaining our idea and results, this paper concerns with the effect of misalignment angle
  (``wobble angle'' in the case of a precessing NS) on the mass measurement precision.  
 \end{abstract}

\section{Introduction}\label{Sec:introduction}

The distribution of masses of neutron stars (NSs) gives insights to understand
their birth mechanisms and evolution histories.
For instance, the mass measurements of the massive pulsar PSR J1614–2230 \cite{2010Natur.467.1081D}
and PSR J0348+0432 \cite{2013Sci...340..448A}
have great impacts on exotic matter physics and studies  of NS interiors.

Currently, mass measurements of NSs are mostly limited to those in
binaries \cite{2011Ap&SS.336...67L}. Since possible mutual mass
transfers in binaries may change the mass distributions of the
component NSs, it is desirable to measure masses of as many isolated NSs as possible.

We have proposed a new method to measure mass of an isolated NS by detecting Coulomb-type phase shifts 
imprinted in gravitational waves (GWs) by the NS gravitational field \cite{2015PhRvD..91h4032O}. 
In that paper, we have shown that the mass of a NS at 1 kpc distance 
and with its ellipticity $10^{-6}$ is typically measurable
with precision of $20\%$ using the third generation GW telescope such
as the Einstein Telescope (ET) \cite{2010CQGra..27s4002P}, 
if the NS is precessing or has a pinned superfluid core and emits GWs with once and twice the spin frequencies.
  
In the following sections, we will review the idea of our method, and
then comment on our assumptions on misalignment angles in
our original paper \cite{2015PhRvD..91h4032O}, namely the misalignment angle of the freely-precessing
NSs and their effect on mass measurement precision.

\section{Method}\label{Sec:method}
Let us consider an isolated NS with mass $M$ located at the coordinate origin
$r=0$. We assume that the NS emits GW with multiple frequencies 
$\omega_{\alpha} = 2\pi f_{\alpha}$($\alpha = 1,2,\cdots$).
The two polarization modes of GWs from the NS can be written as  
\begin{align}
 \bar h_{+} &= \sum_{\alpha}A_{+,\alpha}\cos\Psi_{\alpha},\quad
 \bar h_{\times} = \sum_{\alpha}A_{\times,\alpha}\sin\Psi_{\alpha},\\
 \Psi_{\alpha} &\equiv - \omega_{\alpha} t + \omega_{\alpha} r + \Phi_{\alpha},\\
 \Phi_{\alpha} &\equiv 2\omega_{\alpha}M \ln[2\omega_{\alpha}r] + \phi_{R\alpha}, 
\end{align}
where $\phi_{R\alpha}$ denotes the constant reference GW phases.
The logarithmic term in $\Phi_{\alpha}$ is the Coulomb phase shift due 
to the static part of the NS gravitational field \cite{2015PhRvD..91h4032O,Asada:1997zu}.
All the searches for GWs from rapidly rotating NSs conducted so far 
have not taken into account these Coulomb phase shifts.  

Our method to measure the mass $M$ of a NS and results are as follows \cite{2015PhRvD..91h4032O}.
Suppose that we determine GW phases $\Phi_{\alpha}$ for two or more frequency modes (say, $\omega_{1}$ and 
$\omega_{2}$ with the ratio $K \equiv \omega_{2}/\omega_{1}$). 
If these two modes have such a property that
\begin{align}
 \phi_{R2}-K\phi_{R1} = 0,
 \label{eq:condition_on_phase}
\end{align}
we can determine the mass from 
the following combination, 
\begin{align}
 \Phi_{2} - K \Phi_{1} = 2\omega_{2}M\ln K
 \simeq 0.06
 \left(\frac{M}{1.4M_\odot}\right)
 \left(\frac{f_2}{1{\rm kHz}}\right)
 \left(\frac{\ln K}{\ln2}\right).
 \label{eq:mass_estimator}
\end{align} 
The condition (\ref{eq:condition_on_phase}) is satisfied for 
a freely-precessing isolated bi-axial NS, which may emit 
GWs (approximately) at once and twice the spin frequencies ($K=2$).
Another possibility can be found in the model proposed by Jones \cite{2010MNRAS.402.2503J}.
He showed that if a tri-axial NS contains a pinned superfluid core and none of the axes of the principal
moments of inertia of the solid crust aligns with the spin angular
velocity vector of the core, then such a NS emits GWs at once and twice the
spin frequencies. Notably, the condition (\ref{eq:condition_on_phase}) is hold
when such a NS is bi-axial (but none of the axes of the principal moments of
inertia is aligned with the spin axis).

The GWs from these two models of NSs have (See
\cite{2015PhRvD..91h4032O,Jaranowski:1998qm,2014CQGra..31j5011B,2015MNRAS.453...53J,2015MNRAS.453.4399P})   
\begin{alignat}{2}
 A_{+,1}     &= \dfrac{1}{4} h_{0}\sin 2\theta \sin\iota\cos\iota,
 &\quad
 A_{\times,1}&= \dfrac{1}{4} h_{0} \sin 2\theta \sin\iota,
 \label{Eq:GWamplitude_1}\\ 
 A_{+,2}     &= \dfrac{1}{2} h_{0}\sin^{2}\theta \left( 1+\cos^{2}\iota
 \right),
 &\quad
 A_{\times,2}&= h_{0}\sin^{2}\theta\cos\iota, 
 \label{Eq:GWamplitude_2}  
\end{alignat}
where the inclination angle $\iota$ is defined as the angle between the rotational axis and 
the line-of-sight, and 
the misalignment angle $\theta$ is defined as the angle between the principal axis and the angular momentum axis. 
The overall amplitude is $h_0 = 4\varepsilon I \omega_1^2/r$ where
$I$ denotes the star's average moment of inertia and the ellipticity $\varepsilon$ quantifies
the degree of the non-axisymmetry.

Assuming these models, we have performed Monte-Carlo simulations 
to study the measurement precision of NS mass using Eq. (\ref{eq:mass_estimator}) for three-year ET observation \cite{2015PhRvD..91h4032O}. 
For those simulations, we have randomly selected sets of waveform parameters $\left(\theta, \iota, \phi_R, \psi, \alpha, \delta \right)$ 
where $\phi_R$ is the reference phase, $\psi$ is the GW polarization phase, and $\alpha$ and $\delta$ are the sky position of the NS. 
Uniform distributions are assumed for those parameters.
The NS moment of inertia $I$ and its ellipticity $\varepsilon$ are set to be 
$10^{45}$ g$\cdot$cm$^2$ and $10^{-6}$
\cite{2009PhRvL.102s1102H,2012MNRAS.426.2404H}, respectively.
We adopted several representative values of $r$ (1 kpc, 10 kpc, and 50 kpc) and $f_1$ (300 Hz and 500 Hz). 
Then, We estimate the measurement error in the NS mass for each simulation. 

In the ideal case where all the waveform parameters except for $\phi_{R}$ and $M$ are known, the measurement error in the GW
phase $\Phi_{\alpha}$ is $1/\rho$ where $\rho$ is the signal-to-noise ratio (SNR). 
Hence, GW detections with $\rho \gtrsim 100$ for both modes may suffice to
estimate the NS mass as indicated in Eq. (\ref{eq:mass_estimator}). 
When all the waveform parameters are unknown in advance, the
correlations among the parameters degrade the mass measurement precision.
To improve the mass measurement precision, we have assumed that 
the spin frequency, the spin-down rate, the sky position of the NS 
are known in advance from electromagnetic observations or GW 
observations by the second generation GW detectors such as the advanced LIGO \cite{Abbott:2007kv},
advanced Virgo \cite{Accadia:2012zzb}, and KAGRA  \cite{Aso:2013eba}.

The resulting cumulative distributions of mass measurement precision are plotted in
the Fig. 1 of \cite{2015PhRvD..91h4032O}. For example, we
found that the mass of the NS with its spin frequency 500 Hz and its ellipticity
is typically measurable with an accuracy of 20\% using the ET.

\section{Misalignment angle}\label{Sec:misalignment}

We have assumed a uniform distribution for the misalignment angle
$\theta$ in our Monte Carlo simulations in the previous paper \cite{2015PhRvD..91h4032O}.
In the case of a NS containing a pinned superfluid core, there seems to be neither observational nor theoretical constraint on $\theta$. 
In fact, a uniform distribution for the misalignment angle
$\theta$ is assumed in this model \cite{2014CQGra..31j5011B,2015MNRAS.453...53J,2015MNRAS.453.4399P}.

A possible range of the misalignment angle $\theta$ in the case of
a freely precessing NS is not well-known. 
Two known examples indicate smaller misalignment angles: 
$\theta\simeq 3$ degrees (PSR B1828-11) \cite{2001ApJ...556..392L}
and $\theta\simeq 0.8$
degrees (PSR B1642-03) \cite{2001ApJ...552..321S}. On the other hand, 
the theoretical maximum possible value of $\theta$ is estimated by Eq. (24) of
\cite{2002MNRAS.331..203J} as $\theta_{\rm max} \simeq 10$ degrees $(500
{\rm Hz}/f_1)(u_{\rm break}/{10^{-2}})$ where $u_{\rm break}$ is the
breaking strain of the NS crust. For the 
freely precessing magnetars recently discovered in 
\cite{2014PhRvL.112q1102M,2015PASJ..tmp..263M},
parameter degeneracies prevent from inferring misalignment angles.
In any case, one may find it questionable to 
use a uniform distributions for $\theta$ in the case of a precessing
NS. 

\begin{figure}[htbp] 
\centering 
\includegraphics[width=8cm,clip]{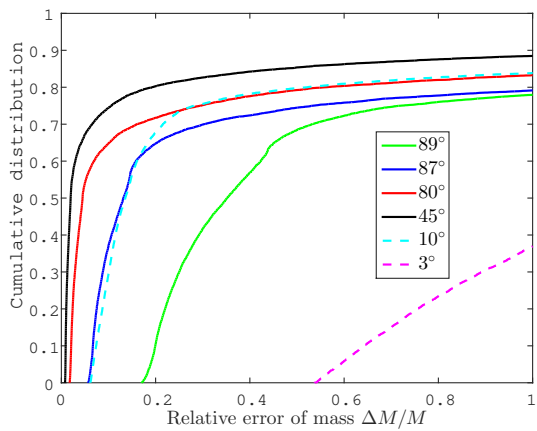}
\caption{\label{Fig:mass_accuracy}
 The cumulative distribution functions for relative
 measurement errors in NS mass $\Delta M_{\text{NS}}/M_{\text{NS}}$. 
 This figure is obtained by Monte Carlo simulations where the NS sky position, 
 polarization angle, reference GW phase, and inclination angles are
 randomly chosen from uniform distributions.
 The distance of $r = 1$ kpc, spin frequency of $f = 500$Hz, and three year 
 observation by the ET are assumed.  
 The solid lines correspond to NSs with the misalignment angles of 
 $\theta = 45^{\circ}$ (black line), $80^{\circ}$ (red line), $87^{\circ}$ (blue line), $89^{\circ}$ (green line), 
 while the dashed lines correspond to $\theta = 10^{\circ}$ (cyan line) and $3^{\circ}$ (magenta line), respectively. 
 While a NS with a pinned superfluid core may have any
 values of $\theta$, a freely precessing NS may be limited by $\theta < 10$ degrees. 
 Mass measurement precision for a freely precessing NS
 may be worse than that for a NS with a pinned superfluid core.
 }
\end{figure}

To see how the misalignment angle affects the measurement
precision of NS mass in our method, we have conducted 10,000 Monte Carlo simulations for three-year ET observations 
similar to the previous paper, but for several fixed misalignment angles.
In these simulations, 
the sky position, polarization angle, reference phase, and inclination angles 
are randomly chosen from uniform distributions.
The distance of $r = 1$ kpc and spin frequency of $f = 500$ Hz are assumed. 
Fisher matrix method is used to estimate measurement errors of the waveform parameters.
As in the previous simulations, the frequency, spin-down rate, sky position of each NS are assumed to be known. 
The results are shown in Fig. \ref{Fig:mass_accuracy}. 
This figure indicates that 
masses of more than 70\% of NSs are measurable with a precision of $\Delta M/M \simeq 0.2$
if $10^{\circ} \leq \theta \leq 80^{\circ}$.
On the other hand, if the misalignment angles are smaller than $10$
degrees, as suggested by PSR B1828-11 and PSR B1642-03, even the third
generation GW telescope cannot determine the NS mass with sufficient
precision (e.g. $\Delta M/M \simeq 0.8$ for 20\% of NSs in the case of $\theta = 3^{\circ}$). 
This is because the amplitudes for both modes become smaller for the smaller misalignment angle 
as indicated by Eqs. (\ref{Eq:GWamplitude_1}) and (\ref{Eq:GWamplitude_2}).

\section{Acknowledgments}
We thank Toshio Nakano for informing us that the misalignment angles
are not determined for the freely precessing magnetars
\cite{2014PhRvL.112q1102M,2015PASJ..tmp..263M} 
due to parameter degeneracy. 
We also thank Bruce Allen for pointing out
that mass measurement precision would be degraded for freely
precessing stars with $\theta$ close to zero. 
This work is supported by JSPS Fellows Grant No. 26.8636 (K. E.),
JSPS Grant-in-Aid for Young Scientists Grant No. 25800126, and the MEXT Grant-in-Aid for
Scientific Research on Innovative Areas (Grant Number 24103005) (Y. I.).

\section*{References}
\bibliography{reference}
   
\end{document}